

\def\singlespace{\normalbaselines}
\def\oneandahalfspace{\baselineskip=1.15\normalbaselineskip plus 1pt
\lineskip=2pt\lineskiplimit=1pt}

\def\itemitemitem{\par\indent\indent \hangindent3\parindent \textindent}

\def\np{\vfill\eject}
\def\nl{\hfil\break}

\def\nofirstpagenoten{\nopagenumbers\footline={\ifnum\pageno>1\tenrm
\hss\folio\hss\fi}}
\def\nofirstpagenotwelve{\nopagenumbers\footline={\ifnum\pageno>1\twelverm
\hss\folio\hss\fi}}
\def\leaderfill{\leaders\hbox to 1em{\hss.\hss}\hfill}
\def\ft#1#2{{\textstyle{{#1}\over{#2}}}}
\def\frac#1/#2{\leavevmode\kern.1em
\raise.5ex\hbox{\the\scriptfont0 #1}\kern-.1em/\kern-.15em
\lower.25ex\hbox{\the\scriptfont0 #2}}
\def\sfrac#1/#2{\leavevmode\kern.1em
\raise.5ex\hbox{\the\scriptscriptfont0 #1}\kern-.1em/\kern-.15em
\lower.25ex\hbox{\the\scriptscriptfont0 #2}}


\parindent=20pt
\def\narrow{\advance\leftskip by 40pt \advance\rightskip by 40pt}

\def\AB{\bigskip
        \centerline{\bf ABSTRACT}\medskip\narrow}
\def\nonarrower{\advance\leftskip by -40pt\advance\rightskip by -40pt}
\def\AE{\bigskip\nonarrower}

\def\boxit#1{\vbox{\hrule\hbox{\vrule\kern3pt
        \vbox{\kern3pt#1\kern3pt}\kern3pt\vrule}\hrule}}

\def\gtorder{\mathrel{\raise.3ex\hbox{$>$}\mkern-14mu
             \lower0.6ex\hbox{$\sim$}}}
\def\ltorder{\mathrel{\raise.3ex\hbox{$<$}|mkern-14mu
             \lower0.6ex\hbox{\sim$}}}
\def\dalemb#1#2{{\vbox{\hrule height .#2pt
        \hbox{\vrule width.#2pt height#1pt \kern#1pt
                \vrule width.#2pt}
        \hrule height.#2pt}}}

\font\fourteentt=cmtt10 scaled \magstep2
\font\fourteenbf=cmbx12 scaled \magstep1
\font\fourteenrm=cmr12 scaled \magstep1
\font\fourteeni=cmmi12 scaled \magstep1
\font\fourteenss=cmss12 scaled \magstep1
\font\fourteensy=cmsy10 scaled \magstep2
\font\fourteensl=cmsl12 scaled \magstep1
\font\fourteenex=cmex10 scaled \magstep2
\font\fourteenit=cmti12 scaled \magstep1
\font\twelvett=cmtt10 scaled \magstep1 \font\twelvebf=cmbx12
\font\twelverm=cmr12 \font\twelvei=cmmi12
\font\twelvess=cmss12 \font\twelvesy=cmsy10 scaled \magstep1
\font\twelvesl=cmsl12 \font\twelveex=cmex10 scaled \magstep1
\font\twelveit=cmti12
\font\tenss=cmss10
 
 \font\ninebf=cmbx7 scaled \magstep1
\font\ninerm=cmr7 scaled \magstep1 \font\ninei=cmmi7 scaled \magstep1
\font\ninesy=cmsy7 scaled \magstep1 
\font\eightrm=cmr7 scaled 1140 
 
\font\sevenbf=cmbx7 \font\sevenrm=cmr7 \font\seveni=cmmi7
\font\sevensy=cmsy7 

\catcode`@=11
\newskip\ttglue
\newfam\ssfam

\def\fourteenpoint{\def\rm{\fam0\fourteenrm}
\textfont0=\fourteenrm \scriptfont0=\tenrm \scriptscriptfont0=\sevenrm
\textfont1=\fourteeni \scriptfont1=\teni \scriptscriptfont1=\seveni
\textfont2=\fourteensy \scriptfont2=\tensy \scriptscriptfont2=\sevensy
\textfont3=\fourteenex \scriptfont3=\fourteenex \scriptscriptfont3=\fourteenex
\def\it{\fam\itfam\fourteenit} \textfont\itfam=\fourteenit
\def\sl{\fam\slfam\fourteensl} \textfont\slfam=\fourteensl
\def\bf{\fam\bffam\fourteenbf} \textfont\bffam=\fourteenbf
\scriptfont\bffam=\tenbf \scriptscriptfont\bffam=\sevenbf
\def\tt{\fam\ttfam\fourteentt} \textfont\ttfam=\fourteentt
\def\ss{\fam\ssfam\fourteenss} \textfont\ssfam=\fourteenss
\tt \ttglue=.5em plus .25em minus .15em
\normalbaselineskip=16pt
\abovedisplayskip=16pt plus 4pt minus 12pt
\belowdisplayskip=16pt plus 4pt minus 12pt
\abovedisplayshortskip=0pt plus 4pt
\belowdisplayshortskip=9pt plus 4pt minus 6pt
\parskip=5pt plus 1.5pt
\setbox\strutbox=\hbox{\vrule height12pt depth5pt width0pt}
\let\sc=\tenrm
\let\big=\fourteenbig \normalbaselines\rm}
\def\fourteenbig#1{{\hbox{$\left#1\vbox to12pt{}\right.\n@space$}}}

\def\twelvepoint{\def\rm{\fam0\twelverm}
\textfont0=\twelverm \scriptfont0=\ninerm \scriptscriptfont0=\sevenrm
\textfont1=\twelvei \scriptfont1=\ninei \scriptscriptfont1=\seveni
\textfont2=\twelvesy \scriptfont2=\ninesy \scriptscriptfont2=\sevensy
\textfont3=\twelveex \scriptfont3=\twelveex \scriptscriptfont3=\twelveex
\def\it{\fam\itfam\twelveit} \textfont\itfam=\twelveit
\def\sl{\fam\slfam\twelvesl} \textfont\slfam=\twelvesl
\def\bf{\fam\bffam\twelvebf} \textfont\bffam=\twelvebf
\scriptfont\bffam=\ninebf \scriptscriptfont\bffam=\sevenbf
\def\tt{\fam\ttfam\twelvett} \textfont\ttfam=\twelvett
\def\ss{\fam\ssfam\twelvess} \textfont\ssfam=\twelvess
\tt \ttglue=.5em plus .25em minus .15em
\normalbaselineskip=14pt
\abovedisplayskip=14pt plus 3pt minus 10pt
\belowdisplayskip=14pt plus 3pt minus 10pt
\abovedisplayshortskip=0pt plus 3pt
\belowdisplayshortskip=8pt plus 3pt minus 5pt
\parskip=3pt plus 1.5pt
\setbox\strutbox=\hbox{\vrule height10pt depth4pt width0pt}
\let\sc=\ninerm
\let\big=\twelvebig \normalbaselines\rm}
\def\twelvebig#1{{\hbox{$\left#1\vbox to10pt{}\right.\n@space$}}}

\def\tenpoint{\def\rm{\fam0\tenrm}
\textfont0=\tenrm \scriptfont0=\sevenrm \scriptscriptfont0=\fiverm
\textfont1=\teni \scriptfont1=\seveni \scriptscriptfont1=\fivei
\textfont2=\tensy \scriptfont2=\sevensy \scriptscriptfont2=\fivesy
\textfont3=\tenex \scriptfont3=\tenex \scriptscriptfont3=\tenex
\def\it{\fam\itfam\tenit} \textfont\itfam=\tenit
\def\sl{\fam\slfam\tensl} \textfont\slfam=\tensl
\def\bf{\fam\bffam\tenbf} \textfont\bffam=\tenbf
\scriptfont\bffam=\sevenbf \scriptscriptfont\bffam=\fivebf
\def\tt{\fam\ttfam\tentt} \textfont\ttfam=\tentt
\def\ss{\fam\ssfam\tenss} \textfont\ssfam=\tenss
\tt \ttglue=.5em plus .25em minus .15em
\normalbaselineskip=12pt
\abovedisplayskip=12pt plus 3pt minus 9pt
\belowdisplayskip=12pt plus 3pt minus 9pt
\abovedisplayshortskip=0pt plus 3pt
\belowdisplayshortskip=7pt plus 3pt minus 4pt
\parskip=0.0pt plus 1.0pt
\setbox\strutbox=\hbox{\vrule height8.5pt depth3.5pt width0pt}
\let\sc=\eightrm
\let\big=\tenbig \normalbaselines\rm}
\def\tenbig#1{{\hbox{$\left#1\vbox to8.5pt{}\right.\n@space$}}}
\let\rawfootnote=\footnote \def\footnote#1#2{{\rm\parskip=0pt\rawfootnote{#1}
{#2\hfill\vrule height 0pt depth 6pt width 0pt}}}

\def\tenfoot{\tenpoint\hskip-\parindent\hskip-.1cm}

\twelvepoint
\def\sbullet{\raise.2em\hbox{$\scriptscriptstyle\bullet$}}
\nofirstpagenotwelve
\hsize=16.5 truecm
\baselineskip 15pt

\def\ft#1#2{{\textstyle{{#1}\over{#2}}}}
\def\sss{\scriptscriptstyle}

\def\im{{\rm i}}
\def\tL{\widetilde L}
\def\Z{\ss \rlap Z\mkern3mu Z}
\def\ph{{\hat p}}

\def\cramp{\medmuskip = -2mu plus 1mu minus 2mu}
\def\uncramp{\medmuskip = 4mu plus 2mu minus 4mu}

\oneandahalfspace
\rightline{CTP TAMU--68/91}
\rightline{Imperial/TP/90-91/40}
\rightline{IC/91/241}
\rightline{September 1991}

\vskip 2truecm
\centerline{\bf The $W_3$ String Spectrum}
\vskip 1.5truecm
\centerline{C.N. Pope,$^
1$\footnote{$^
\star$}{\tenfoot Supported in part by the
U.S. Department of Energy, under
grant DE-FG05-91ER40633.}
L.J. Romans,$^
2$\footnote{$^
{\ddag}$}{\tenfoot
Supported by a National Research Council-NASA/JPL
Research Associateship.}
E. Sezgin,$^
1$\footnote{$^
\$ $}{\tenfoot Supported in part by the
National Science Foundation, under
grant PHY-9106593.} and K.S. Stelle$^
3$}
\vskip 1.5truecm
\itemitemitem{$^
1$}{\it Center
for Theoretical Physics,
Texas A\&M University, College Station\nl TX 77843--4242, USA\/}
\itemitemitem{$^
2$}{\it Jet Propulsion Laboratory 301-150, California Institute of Technology,
\nl Pasadena, CA 91109, USA\/}
\itemitemitem{$^
3$}{\it The
Blackett Laboratory, Imperial College, London SW7 2BZ\/}
\vskip 1.5truecm
\AB\singlespace
      We study the spectrum of $W_3$ strings.  In particular, we show that
for appropriately chosen space-time signature, one of the scalar fields is
singled out by the spin-3 constraint and is ``frozen'':  no creation operators
from it can appear in physical states and the corresponding momentum must
assume a specific fixed value.  The remaining theory is unitary and resembles
an ordinary string theory in $d\ne26$ with anomalies cancelled by appropriate
background charges.  In the case of the $W_3$ string, however, the spin-two
``graviton'' is massive.
\AE\oneandahalfspace
\np
\noindent
{\bf 1. Introduction}
\bigskip

     The theory of $W_3$ strings starts from the construction of a
non-chiral anomaly-free quantum $W_3$ gravity theory [1].  This is
achieved [2] by taking a quantum realisation of the $W_3$ algebra with
central charge $c=100$ [3].  Such realisations can be found for two [4]
or more [5] scalar fields.  In all cases, including that of $n=100$
scalar fields, these realisations require background charges for some
of the scalar fields [5].  In the $W_3$ string theory, the $n$ scalar
fields may be interpreted as spacetime coordinates.  At the classical
level, the equations of motion for the spin-2 and spin-3 gauge fields
of $W_3$ gravity impose the vanishing of the left-moving and
right-moving spin-2 and spin-3 matter currents.  The realisations of
the $W_3$ algebra found in [5] have the property that one scalar field,
which we shall denote by $\varphi_1$, is distinguished in the way that
it appears in the spin-3 current; the other $n-1$ fields all appear
only through their stress tensor.  Upon use of the spin-2 constraints,
this has the consequence of ``freezing'' the $\varphi_1$ coordinate
[1]; {\it i.e.}\ the spin-3 constraints imply $(\partial\varphi_1)^
3
=(\bar\partial\varphi_1)^
3=0$, and hence $\varphi_1$ is constant.  In
the present paper, we shall extend this classical picture of $W_3$
strings to the quantum level by discussing the quantum version of the
``coordinate-freezing'' phenomenon.  We shall also discuss the
spectrum of physical states.  Although we shall mostly be considering
closed $W_3$ strings, our discussion generalises straightforwardly to
the case of open $W_3$ strings.  Some preliminary quantum results have
already been given in [1].

     We shall denote the $n$ scalars  by ($\varphi_1$, $\varphi_2$,
$X^
\mu$), with $\mu=0,3,4,\ldots,n-1$.  In order to have a
Minkowski-signature spacetime, we shall take $X^
0$ to be timelike and
all the remaining coordinates to be
spacelike.\footnote{$^
\ast$}{\tenfoot Note that we have made a
different choice of signature for the $\varphi_1$ coordinate from that
in [1].  This is in order to ensure the unitarity of the theory, as we
shall see later.}  The left-moving matter currents take the forms
$$\eqalignno{
T_{\rm mat}&= {\cal
T}-\ft12(\partial\varphi_1)^
2-\ft12(\partial\varphi_2)^
2-
(Q_1\partial^
2\varphi_1+Q_2\partial^
2\varphi_2)&(1a)\cr
W_{\rm mat}&= -{2\im \over\sqrt{261}}\Big\{\ft13(\partial\varphi_1)^
3-
\partial\varphi_1(\partial\varphi_2)^
2+(Q_1\partial\varphi_1
\partial^
2\varphi_1-2Q_2\partial\varphi_1\partial^
2\varphi_2-Q_1
\partial\varphi_2\partial^
2\varphi_2)\cr
&\phantom{=-
{2\over\sqrt{261}}\Big\{}+(\ft13Q_1^
2\partial^
3\varphi_1
-Q_1Q_2\partial^
3\varphi_2)+2\partial\varphi_1{\cal
T}+Q_1\partial{\cal T}\Big\},&(1b)\cr}
$$ where
$\cal T$ is the stress tensor for the $D=n-2$ free scalar fields
$X^
\mu$ without background charges:
$$
{\cal T}=
-\ft12\eta_{\mu\nu}\partial X^
\mu\partial X^
\nu.\eqno(2)
$$
The right-moving currents $\widetilde T_{\rm mat}$ and $\widetilde
W_{\rm mat}$ are defined similarly, with $\partial$ replaced by
$\bar\partial$.  The background charges
$Q_1$ and $Q_2$ for $\varphi_1$ and $\varphi_2$ are given
by [2]
$$\eqalign{
Q_1^
2&=\ft{49}8\cr
Q_2^
2&=\ft1{12}(\ft{49}2-D),\cr}\eqno(3)
$$
and are needed to obtain a total matter central charge $c_{\rm
mat}=100$.  Initially, we shall assume that $D\le24$, so that $Q_2$ is
real.

     We expand the coordinates in Laurent series as usual:
$$\eqalign{
\im\partial\varphi_1&=\sum_m\alpha^
1_mz^
{-m-1}\cr
\im\partial\varphi_2&=\sum_m\alpha^
2_mz^
{-m-1}\cr
\im\partial X^
\mu&=\sum_m\alpha^
\mu_mz^
{-m-1}.\cr}\eqno(4)
$$
Upon quantisation, the oscillators $\alpha_m$ satisfy the commutation
relations
$$
[\alpha^
i_m,\alpha^
j_n]=m\eta^
{ij}\delta_{m+n,0}\qquad\qquad i=1,2,\mu
\eqno(5)
$$
where $\eta^
{00}=-1$ and all the other diagonal elements equal $+1$.
Expanding the spin-2 current $T_{\rm mat}=\sum_nL_nz^
{-n-2}$ and
the spin-3 current $W_{\rm mat}=\sum_nW_nz^
{-n-3}$ in terms of the
oscillators, we have
$$\eqalignno{
L_n&=\ft12\sum_m : \Big(\alpha^
1_{n-m}\alpha^
1_m +
\alpha^
2_{n-m}\alpha^
2_m +
\eta_{\mu\nu}\alpha^
\mu_{n-m}\alpha^
\nu_m\Big):
- \im(n+1)\Big(Q_1\alpha^
1_n + Q_2\alpha^
2_n\Big)&(6)\cr
W_n&={2\over\sqrt{261}}\Big\{\sum_{p,q}
:\Big(\ft13\alpha^
1_{n-p-q}\alpha^
1_p\alpha^
1_q -
\alpha^
1_{n-p-q}\alpha^
2_p\alpha^
2_q -
\alpha^
1_{n-p-q}\alpha^
\mu_p\alpha^
\nu_q\eta_{\mu\nu}\Big):\cr
&\phantom{={2\over\sqrt2}}
+ \im\sum_m(m+1):\Big(-Q_1\alpha^
1_{n-m}\alpha^
1_m +
2Q_2\alpha^
1_{n-m}\alpha^
2_m + Q_1\alpha^
2_{n-m}\alpha^
2_m +
Q_1\alpha^
\mu_{n-m}\alpha^
\nu_m\eta_{\mu\nu}\Big):\cr
&\phantom{={2\over\sqrt2}}
+ (n+1)(n+2)\Big(-\ft13Q_1^
2\alpha^
1_n +
Q_1Q_2\alpha^
2_n\Big)\Big\}&(7)\cr}
$$
The hermiticity conditions $L^
\dagger_n=L_{-n}$ and $W^
\dagger_n
=W_{-n}$ imply that ${\alpha^
i_n}^
\dagger=\alpha^
i_{-n}$,
($i=1,2,\mu$), except for $\alpha^
1_0$ and $\alpha^
2_0$, which must
satisfy conditions that are modified by the background charges:
$$
{\alpha^
1_0}^
\dagger=\alpha^
1_0-2 \im Q_1,\qquad {\alpha^
2_0}^
\dagger=
\alpha^
2_0 -2\im Q_2.\eqno(8)
$$
For these, it is convenient to introduce shifted oscillators
${\hat\alpha}^
1_0$ and ${\hat\alpha}^
2_0$ that are Hermitean:
$$
{\hat\alpha}^
1_0=\alpha^
1_0-\im Q_1,\qquad
{\hat\alpha}^
2_0=\alpha^
2_0-\im Q_1.\eqno(9)
$$

     As in ordinary string theory, the quantum constraints that are to
be imposed on physical states are those corresponding to the
Laurent-mode expansion coefficients for the currents with non-negative
indices.  It is sufficient to impose just the constraints for $L_0$,
$L_1$, $L_2$ and $W_0$ (together with their right-moving counterparts),
since all the rest follow by commutation.  It is known from the
BRST-nilpotency conditions derived in [3] that the intercepts for
$L_0$ and $W_0$ are $-4$ and 0 respectively, so the
constraints to be imposed on physical states are [1]:
$$
\eqalignno{
L_0-4&=0,\quad L_1=0,\quad L_2=0,\quad W_0=0,&(10a)\cr
\tL_0-4&=0,\quad \tL_1=0,\quad \tL_2=0,\quad \widetilde
W_0=0.&(10b)\cr}
$$
It is convenient to make use of the $L_n$ constraints to simplify
$W_0$.  After imposing the $L_n$ constraints in (10$a$), we find that
$W_0$ can be rewritten as
$$
\eqalign{
W_0&={1\over 3\sqrt{261}}\Big\{ 5{\sum}' :\alpha^
1_{-p-q}
\alpha^
1_p\alpha^
1_q : + {\hat\alpha}^
1_0 \Big( 8({\hat\alpha}^
1_0)^
2
+36 {\cal N}^
{(1)} +1 \Big)\cr
&\phantom{{1\over 3\sqrt{261}}\Big\{}-12\im Q_1 \sum_{n>0} n
\alpha^
1_{-n} \alpha^
1_n -12 \sum_{n>0} {\hat L}_{-n} \alpha^
1_n
\Big\},\cr}\eqno(11)
$$
where ${\hat\alpha}^
1_0$ is given by (9), ${\cal N}^
{(1)}$ denotes the
number operator for $\alpha^
1$ oscillators
$$
{\cal N}^
{(1)}=\sum_{n >0} \alpha^
1_{-n} \alpha^
1_n,\eqno(12)
$$
and
${\hat L}_n$ denotes the terms in $L_n$ that do not involve the
$\alpha^
1_m$ oscillators,
$$
{\hat L}_{n}=\ft12\sum_m : \Big(\alpha^
2_{n-m}\alpha^
2_m +
\eta_{\mu\nu}\alpha^
\mu_{n-m}\alpha^
\nu_m\Big):
- \im(n+1) Q_2\alpha^
2_n.\eqno(13)
$$
The prime on the summation symbol in the first term in (11) indicates
that only terms with all Laurent indices non-zero are included.

     To construct the physical states, we begin by defining the Fock
vacuum $\big|{\rm vac}\big\rangle$:
$$
\alpha^
i_n \big|{\rm vac}\big\rangle =0,\qquad\qquad i=1,2,\mu; \qquad
n\ge 0. \eqno(14)
$$
The ``tachyon'' state $\big|p\big\rangle$ is given by
$$
\big|p\big\rangle= \exp \Big(\im p_1 \varphi^
1(0) +\im p_2\varphi^
2(0)
 +\im p_\mu X^
\mu(0) \Big) \big|{\rm vac}\big\rangle,\eqno(15)
$$
and higher states are built by acting on (15) with linear
combinations of products of the creation operators $\alpha^
i_n$,
$n<0$.  Of course, as usual in closed string theory, there is a
global constraint that the left-moving and right-moving level numbers
must be equal.  With this understood, we shall concentrate in the
following on just the left-moving sector.
\np
\noindent
{\bf 2. Freezing the $\varphi^
1$ coordinate}
\bigskip

     We now argue that no physical states can involve any
$\alpha^
1_n$ creation operators.  This is the quantum version of the
``coordinate-freezing'' phenomenon mentioned above at the classical
level [1].  Consider imposing the $W_0$ constraint, with $W_0$ given
by (11), on a general level-$N$ state in the theory.  (The level $N$ is
the sum of the Laurent-mode indices in each monomial of creation
operators forming the state.)  Amongst the monomials at level $N$
will be ones that involve only $\alpha^
1$ oscillators.  From the
form of (11), we see that all terms except the last one preserve the
$\alpha^
1$ level number $N^
{(1)}$, whilst the last term lowers it.
After applying $W_0$ to the state, we may
focus on the terms that still have $\alpha^
1$ level $N^
{(1)}=N$. (The
final term in (11) may thus be neglected in this discussion.)  The
coefficient of each such independent monomial must
vanish, by virtue of the $W_0=0$ constraint.  This gives as many
homogeneous equations as there are coefficients to be determined.  As
we shall see in examples below, these systems are non-degenerate, and
hence the coefficients of all the monomials with $\alpha^
1$ level
equal to $N$ must vanish.  The argument can now be repeated at
level $N^
{(1)}=N-1$ in $\alpha^
1$ oscillators, so that we have just one
$\alpha^
{j\ne 1}_1$ oscillator in each monomial.  Again, we focus on
terms of highest $\alpha^
1$ level, now equal to $N-1$, after applying
$W_0$.  (As before, the last term in (11) may
be neglected.)  The $\alpha^
{j\ne 1}_1$ oscillators commute with the
relevant terms in (11), and again we obtain a non-degenerate set of
homogeneous equations for the monomial coefficients.
Proceeding iteratively, we find that all monomials containing
$\alpha^
1$ creation operators must have zero coefficients.

     Let us illustrate the above discussion with some examples.
At level 1, the only term involving $\alpha^
1$ is of the form
$$
\lambda \alpha^
1_{-1}  \big|p\big\rangle.\eqno(16)
$$
(Since we shall focus on the terms involving a surviving
$\alpha^
1_{-1}$ after applying $W_0$, and since no terms in $W_0$ can
increase the $\alpha^
1$ level $N^
{(1)}$, we do not need to consider the
other possible monomials in a level-1 state.)  Thus applying $W_0$, we
find that the coefficient of $\alpha^
1_{-1}$ is proportional to
$$
  \lambda \Big(\ph_1\big( 8 (\ph_1)^
2+37\big) -12\im Q_1\Big),
\eqno(17)
$$
where $\ph_1$ is the eigenvalue of the Hermitean operator
${\hat\alpha}^
1_0$ defined in (9).  Since $\ph_1$ is real, it follows
from the $W_0$ constraint that $\lambda$ is zero.  Thus, at level 1
there can be no $\alpha^
1$ creation operators in physical states.

     At level 2, the monomials with $\alpha^
1$ level 2 have the form
$$
\Big(\lambda \alpha^
1_{-2} +\mu \alpha^
1_{-1}\alpha^
1_{-1} \Big)
 \big|p\big\rangle.\eqno(18)
$$
After applying $W_0$, the coefficients of the monomials $\alpha^
1_{-2}$
and  $\alpha^
1_{-1}\alpha^
1_{-1}$ must independently vanish.  This
gives the set of equations
$$
\pmatrix{F_2-48\im Q_1&30\cr
30&F_2-24 \im Q_1\cr}
\pmatrix{\lambda\cr\mu}=0,\eqno(19)
$$
where $F_{N^
{(1)}}$ is the real quantity
$$
F_{N^
{(1)}}=\ph_1\Big( 8(\ph_1)^
2+36 N^
{(1)}+1\Big).\eqno(20)
$$
The condition for (19) to admit non-zero solutions for
$\lambda$ and $\mu$ ({\it i.e.}\ the condition that the determinant
of the matrix of coefficients vanish) implies that
$$
F_2=(36\im \pm \ft{12}7) Q_1,\eqno(21)
$$
which is impossible since $F_2$ is real.  Thus $\lambda$ and $\mu$ must
be zero.  This shows that at level 2, physical states can be at most
of level 1 in $\alpha^
1$ oscillators.  But our previous discussion
for level 1 states now carries over without modification to show that
the coefficients of these monomials must also be zero.  Therefore no
$\alpha^
1$ creation operators can appear in level-2 physical states.

     At higher levels the above pattern repeats.  We have explicitly
checked up to level 5 that the $W_0$ constraint produces a
non-degenerate set of homogeneous equations for the coefficients of all
the possible monomials involving $\alpha^
1$ oscillators, showing that
at least up to this level, no $\alpha^
1$ creation operators can
occur in physical states.  The reason is always that the roots of
the various vanishing-determinant conditions would require complex
solutions for the real quantities $F_{N^
{(1)}}$.  The pattern of
the imaginary parts becomes clear from our sequence of low-level
examples.  The imaginary parts $y_r$, $r=1,\ldots,P_{N^
{(1)}}$ of the
roots (where $P_{N^
{(1)}}$ is the number of possible $\alpha^
1$
monomials at level $N^
{(1)}$) turn out in all cases to be given by the
formula $$
y_r=\ft12\Big(u_r+u^
{\phantom{\chi}}_{(P_{N^
{(1)}}+1-r)}\Big),\qquad
r=1,\ldots,P_{N^
{(1)}}.\eqno(22)
$$
The $u_r$ in (22) are the imaginary
parts of the diagonal entries in the matrix of equation coefficients,
organised in monotonically decreasing value.  Thus, $u_r$ is the
eigenvalue of the diagonal operator
$12Q_1\sum_{n>0}n\alpha^
1_{-n}\alpha^
1_n$ corresponding to the
eigenvector created by the $r$'th monomial. Since all the $u_r$ are
positive, it follows that the $y_r$ are all non-zero.  Consequently, the
homogeneous equations for the $\alpha^
1$ monomial coefficients are
non-degenerate and so all these coefficients must vanish.

     As a final illustration of these ideas, we give the
$7\times7$ matrix of equation coefficients for the level-5 monomials
$\alpha^
1_{-5}$, $\alpha^
1_{-4}\alpha^
1_{-1}$,
$\alpha^
1_{-3}\alpha^
1_{-2}$, $\alpha^
1_{-3}(\alpha^
1_{-1})^
2$,
$(\alpha^
1_{-2})^
2\alpha^
1_{-1}$, $\alpha^
1_{-2}(\alpha^
1_{-1})^
3$ and
$(\alpha^
1_{-1})^
5$:
\cramp
$$
\pmatrix{
F_5-300\im Q_1&120&180&0&0&0&0\cr
150&F_5-204\im Q_1&0&180&120&0&0\cr
150&0&F_5-156\im Q_1&30&120&0&0\cr
0&120&30&F_5-132\im Q_1&0&180&0\cr
0&60&90&0&F_5-108\im Q_1&90&0\cr
0&0&0&90&60&F_5-84\im Q_1&300\cr
0&0&0&0&0&30&F_5-60\im Q_1\cr}\eqno(23)
$$
\uncramp
It is straightforward to show that the characteristic equation gives
roots whose imaginary parts are in accordance with (22),
$$
F_5=132\im Q_1,\ (180\im\pm\ft{120}7)Q_1,\ (144\im\pm\ft{60}7)Q_1,\
(132\im\pm\ft{24}7)Q_1.\eqno(24)
$$
Thus, there are no real solutions for $F_5$.

     Having seen that the physical states involve no $\alpha^
1$
oscillators, the remaining content of the $W_0$ constraint on physical
states is, from (11),
$$
\ph_1\Big(8(\ph_1)^
2+1\Big)=0.\eqno(25)
$$
Thus, for all physical states $\ph_1$ is frozen:
$$
\ph_1=0.\eqno(26)
$$
\bigskip
\noindent
{\bf 3. The construction of physical states}
\bigskip

     We have seen from the discussion in the previous section that the
$W_0$ constraint simply has the effect of eliminating the $\varphi^
1$
coordinate from the theory, and freezing the $\ph_1$ momentum as in
(26).  In terms of the original unshifted momentum component $p_1$
(the eigenvalue of the non-hermitean operator $\alpha^
1_0$), this
implies that it simply takes the fixed imaginary value $p_1=\im Q_1$.
The $W_n$ constraints then play no further r\^
ole in the theory.  The
construction of physical states now closely parallels that for string
theory with a background charge.

     Although we have chosen the background charge $Q_2$ to lie in the
$\varphi_2$ direction, we could in fact equally well allow it to lie
in some unspecified direction in the $(D+1)$-dimensional spacetime
spanned by the coordinates $\varphi_2$ and $X^
\mu$, as long as
the (vector) background-charge parameter $Q^
a$, $a=2,\mu$, has the same
magnitude as $Q_2$, {\it i.e.}
$$
Q^
a Q_a=\ft1{12}(\ft{49}2-D).\eqno(27)
$$
It is sometimes advantageous to do this, since this enables us to write the
constraints and physical states in a formally $(D+1)$-dimensional
Lorentz-covariant fashion.  Note, however, that since the constant
vector $Q^
a$ singles out a direction in spacetime,
$(D+1)$-dimensional Lorentz invariance is broken. Thus, bearing in mind
that physical states do not involve $\alpha^
1$ creation operators, we
may rewrite the $L_n$ constraint operators, for application onto such
states, as
$$
\eqalignno{
L_0-4&=\ft12{\hat\alpha}^
a_0{\hat\alpha}^
b_0 \eta_{ab} +
{\cal N}-4 +\ft12 Q_1^
2 +\ft12 Q^
a Q_a,\cr
&=\ft12 {\hat\alpha}^
a_0{\hat\alpha}^
b_0 \eta_{ab} +
{\cal N} +\ft1{12}(2-D),&(28a)\cr
\cr
L_n&=({\hat\alpha}_{0a}-\im n Q_a)\alpha^
a_n + \ft12\eta_{ab}{\sum}'
:\alpha^
a_{n-m} \alpha^
b_m:,\qquad n>0.&(28b)\cr}
$$
The (hermitean) operators ${\hat\alpha}^
a_0$ are related to $\alpha^
a_0$ by
$$
{\hat\alpha}^
a_0=\alpha^
a_0-\im Q^
a;\eqno(29)
$$
${\cal N}$ is the number operator for the $\alpha^
a$ oscillators,
$$
{\cal N}=\sum_{m>0} \eta_{ab}\alpha^
a_{-m} \alpha^
b_m;\eqno(30)
$$
and the prime on the summation in (28$b$) indicates that only terms
with  non-zero Laurent indices are to be included.  Note that the
original momentum components $p^
a$ have a fixed imaginary part $Q^
a$.

     Physical states are built up from the Fock vacuum by applying
linear combinations of monomials in the creation operators $\alpha^
a$
that are homogeneous in the overall level number $N$ (the eigenvalue
of ${\cal N}$). As we remarked earlier, in closed-string theory one
has a global constraint that the left-moving and right-moving level
numbers $N$ and ${\widetilde N}$ must be equal.  It is to be
understood in what follows that we are focussing on just the
left-moving sector of such physical states.

     In this section, we shall build the first few levels of
physical states.  The lowest-lying state, at level $N=0$, is the
``tachyon,'' given by (15) with $p_1=\im Q_1$.  The only
remaining non-trivial constraint in this case is the mass-shell
condition $(L_0-4)\big| p\big\rangle=0$, where $L_0-4$ is given by
(28$a$).  Thus we have\footnote{$^
{\ast}$}{\tenfoot Similar discussions
have been given in [6] for the case of a string in $d>26$ dimensions
with a background charge in the time direction.}
$$
-\ph^
a\ph_a=\ft1{12}(2-D).\eqno(31)
$$
We shall postpone discussing the issue of masslessness until
the next section.  For now, we just remind the reader that since there
is a background charge in the $(D+1)$-dimensional spacetime, which
breaks the $(D+1)$-dimensional Lorentz covariance, the notion of mass is
modified.  The Fock vacuum, by definition, will be taken to have
positive norm.  This implies that the tachyon will also have positive
norm.  Note that, because of the presence of the background charge,
the inner product for tachyon states reads
$$
\eqalign{
\big\langle p'\big| p\big\rangle&=\delta(p'+p-2\im Q)\cr
&=\delta({\bar p}'+p).\cr}\eqno(32)
$$

     At the next level, $N={\widetilde N}=1$, the left-moving sector of
a physical state takes the form
$$
\big|\xi,p\big\rangle = \xi_a \alpha^
a_{-1}\big|p\big\rangle.\eqno(33)
$$
The $L_0$ and $L_1$ constraints are non-trivial, giving, respectively,
$$
-\ph^
a\ph_a=\ft1{12}(26-D)\eqno(34)
$$
and
$$
\big(\ph^
a-\im Q^
a\big)\xi_a=0.\eqno(35)
$$
Since $\ph^
a$ is the eigenvalue of the hermitean operator
${\hat\alpha}^
a_0$ given in (29), (35) may be written as
$$
{\bar p}^
a \xi_a=0.\eqno(36)
$$
The norm of the state (33) is a positive number times
$$
S\equiv\eta^
{ab}{\bar\xi}_a \xi_b.\eqno(37)
$$
We may use (36) to solve for the timelike component $\xi_0$, and
rewrite (37) as
$$
S={\bar\xi}_{\sss A}\xi_{\sss B}M^
{\sss AB}\equiv {\bar\xi}_{\sss
A}\xi_{\sss B} \Big(\delta^
{\sss AB}-{p^
{\sss A} {\bar
p}^
{\sss B}\over |p^
0|^
2}\Big),\eqno(38)
$$
where the indices ${\scriptstyle A,B}\ldots$ run over the spatial
directions $2,3,\ldots,D+1$. Clearly, the  matrix $M^
{\sss AB}$ has
$D$ unit eigenvalues (corresponding to the $D$ eigenvectors
$V_{\sss A}$ satisfying ${\bar p}^
{\sss A}V_{\sss A}=0$), and one
special eigenvalue, which corresponds to the eigenvector $p_{\sss A}$.
On using the mass-shell condition (34), which can be rewritten using
(27) and (29) as $-{\bar p}^
a p_a=\ft18$, we see that this eigenvalue
is equal to
$$
{1\over 8 |p_0|^
2}.\eqno(39)
$$
Thus all the states in (33) have positive norm.  This contrasts with
the situation in ordinary string theory, for which the special
eigenvalue analogous to (39) would be zero.  We shall discuss this
further in the next section.

     At level 2 in the left-moving sector, physical states will have
the form
$$
\big|\beta,\epsilon,p\big\rangle= \Big(\beta_{ab}\alpha^
a_{-1}
\alpha^
b_{-1} +\epsilon_a \alpha^
a_{-2}\Big)\big| p
\big\rangle.\eqno(40)
$$
The conditions following from the $L_0$, $L_1$ and $L_2$ constraints
at this level are
$$
\eqalignno{
-\ph^
a\ph_a&=\ft1{12}(50-D),&(41a)\cr
\beta_{ab}(\ph^
b-\im Q^
b) +\epsilon_a&=0,&(41b)\cr
2(\ph^
a-2\im Q^
a)\epsilon_a +\beta_a{}^
a&=0.&(41c)\cr}
$$
Using (27) and (29), these may be rewritten as
$$
\eqalignno{
-{\bar p}^
ap_a&=\ft{17}8,&(42a)\cr
\beta_{ab}{\bar p}^
b+\epsilon_a&=0,&(42b)\cr
2({\bar p}^
a-\im Q^
a)\epsilon_a +\beta_a{}^
a&=0.&(42c)\cr}
$$
We can use (42$b$) to solve for $\epsilon_a$, giving
$\epsilon_a=-\beta_{ab}{\bar p}^
b$.  From (42$c$), $\beta_{ab}$ must
then satisfy the constraint $\beta_a{}^
a=2{\bar p}^
a ({\bar p}^
b-\im
Q^
b) \beta_{ab}$.

     It is useful to decompose $\beta_{ab}$ in the form
$$
\beta_{ab}=\lambda\eta_{ab}+\mu p_a p_b +\beta^
{\rm T}_{(
a}p_{b)} +\beta^
{\rm TT}_{ab},\eqno(43)
$$
where
$$
{\bar p}^
a\beta^
{\rm T}_a=0,\qquad {\bar p}^
a \beta^
{\rm
TT}_{ab}=0,\qquad \beta^
{\rm TT}_a{}^
a=0.\eqno(44)
$$
Substituting this into the constraint on $\beta_{ab}$ derived from
(42$b$,$c$), one finds that
$$
{\im\over8}\beta^
{\rm T}_a Q^
a=(2\lambda-\ft34\mu)(\ft{119}8+3\im p_a Q^
a).
\eqno(45)
$$
 Using this, and inserting (42$b$)
and (43) into the expression for the norm, which is a positive multiple of
$S\equiv{\bar \epsilon}_a\epsilon^
a +{\bar\beta}_{ab}\beta^
{ab}$, gives
$$
S=\Big(\ft{17}8\Big)^
2
\Big[-8|\lambda|^
2-\ft98|\mu|^
2+3\lambda\bar\mu
+3\bar\lambda \mu\Big]+\ft{17}{256}{\bar \beta}^
{\rm T}_a\beta^
{{\rm T}a}+
{\bar \beta}^
{\rm TT}_{ab}\beta^
{{\rm TT}ab}.\eqno(46)
$$

     The constraint (45) implies that $\beta^
{\rm T}_a$ must be non-zero when
$\mu\ne \ft83\lambda$, so the quadratic form (46) still has a mixing between
the $\lambda$, $\mu$ and $\beta^
{\rm T}_a$ modes.  In fact, it is just the
single mode in $\beta^
{\rm T}_a$ that is parallel to $Q_a$ that mixes with
$\lambda$ and $\mu$.  Thus we must diagonalise the corresponding
three-dimensional quadratic form.  We do this by noting that we may, without
loss of generality, choose $Q_a$ to lie in the $\varphi_2$ direction, and
furthermore we may use the remaining $(D-1)$-dimensional rotation group for the
coordinates $(X^
3,\ldots, X^
{D+1})$ to choose $p_a$ such that only $p_0$, $p_2$
and $p_3$ are non-zero.  Solving for $\beta^
{\rm T}_0$ from (44), and for
$\beta^
{\rm T}_2$ from (45), we obtain the three-dimensional contribution $S_3$
to $S$, sesquilinear in $(\lambda,\mu,\beta^
{\rm T}_3)$:
$$
\eqalign{
S_3&=Q_2^
{-2}\Big(\ft{17}8+p_3^
2\Big)\Big|6\mu-16\lambda\Big|^
2
\Big[\big(\ft{119}8\big)^
2 +9 Q_2^
2|p_2|^
2 -\ft{357}4 Q^
2\Big]\cr
&+\Big\{\im Q_2^
{-1} \beta^
{\rm T}_3 (6\bar\mu-16\bar \lambda)(\ft{119}8-3\im
\bar p_2 Q_2)p_2 p_3 +\hbox{ c.c.} \Big\}\cr
&+(\ft{17}8+|p_2|^
2)|\beta^
{\rm T}_3|^
2 -17p_0^
2\Big( 32|\lambda|^
2 +\ft{9}2
|\mu|^
2-12\lambda \bar\mu -12\bar\lambda \mu\Big).\cr}\eqno(47)
$$
(Note that since $Q_a$ has been chosen to lie along the 2 direction, the
momentum components $p_a$ in all other directions are real.)  It is
straightforward to see that the quadratic form (47) has eigenvalues $\sigma_1$,
$\sigma_2$ and $\sigma_3$ satisfying
$$\eqalignno{
\sigma_1&=0,&(48a)\cr
\sigma_2\sigma_3&={1241\over 128 Q_2^
2}p_0^
2\Big(14161-5848Q_2^
2+512 Q_2^
4 +512
Q_2^
2 \ph_2^
2\Big),&(48b)\cr
\sigma_2+\sigma_3&={1\over 128 Q_2^
2}\Big(17573801-7257096Q_2^
2 +635520 Q_2^
4
+8270024 p_3^
2 -3415232 Q_2^
2 p_3^
2\cr
&\qquad\qquad +336384 Q_2^
4 p_3^
2 +635520 Q_2^
2 \ph_2^
2 +336384 Q_2^
2 p_3^
2
\ph_2^
2 \Big). &(48c)\cr}
$$
 From (3), we see that $Q_2^
2$ must lie in the range $\ft1{24}\le Q_2^
2 \le
\ft{49}{24}$, and it is easy to verify that the expressions (48$b$) and (48$c$)
are always positive for $Q_2^
2$ in this range.  Thus $S_3$ represents the
contribution of one zero-norm state and two positive-norm states in the level-2
physical spectrum.

To analyse the remaining states, one may use the
expressions in (44) to solve for the components of $\beta^
{\rm T}_a$ and
$\beta^
{\rm TT}_{ab}$ with timelike indices in terms of $\beta^
{\rm T}_{\sss A}$
and $\beta^
{\rm TT}_{\sss AB}$.  The eigenvalues of the resulting quadratic
form may then be evaluated in a similar fashion to our level-1
calculation.  One finds that the ${\bar \beta}^
{\rm T}_a\beta^
{{\rm T}a}$ and
${\bar \beta}^
{\rm TT}_{ab}\beta^
{{\rm TT}ab}$ terms in (46) have
positive-definite norm.  It should be emphasised that our
results for the norms of these states are not sensitive to the value of
$D$ (except that up until now we have assumed that $D\le24$ in order for
$Q_2$ to be real).  This dimension independence arises because the
background charges ensure that we have a critical theory.

     The restriction $D\le24$ that we have imposed up until now in
order to keep $Q_a$ real may simply be relaxed by analytically
continuing our results now to general $D$.  All the above norm
calculations give $D$-independent results and remain unchanged.
For $D\ge25$, the analogue of the inner product relation (32) is
$$
\big\langle p'\big| p\big\rangle=\delta(p'+p-2B),\eqno(49)
$$
where $B_a=\im Q_a$ is real.  Thus, momentum conservation requires the
same sort of shifts in $p^
a$  in norm calculations as those that were
obtained above in the $D\le24$ region under complex conjugation
$p_a\to\bar p_a=p_a-2\im Q_a$.  Consequently, our above results extend
straightforwardly to the $D\ge25$ region.
\bigskip
\noindent
{\bf 4. Masses and compactification}
\bigskip

     We have seen above that the level-1 states of the $W_3$ string
differ from those of the ordinary bosonic string in that there is no
zero-norm state.  This is an indication that the physical states at
this level are {\it not} massless.  What then is one to make of the
mass-shell condition (34), which seems to indicate the possibility of
massless states? (This question has already been raised in [1].)  Since
the presence of the background charge (27) breaks Lorentz covariance
in $D+1$ dimensions, there is not a clear meaning to ``mass'' as a
Lorentz Casimir operator.  In order to clarify this issue, we shall
now consider compactifying the dimension associated with the
background charge and shall discuss the resulting Lorentz-covariant
mass formula in the remaining $D$ dimensions.  For these purposes, it
is convenient to revert to having a definite choice of the direction
associated to the background charge and we shall again let this be the
$\varphi_2$ direction.

     In order to compactify $\varphi_2$, {\it i.e.}\ to let it take
its values in a circle $S^
1$, we have to ensure that there is a
symmetry of the quantum path integral under some shift
$\Delta$ in the value of $\varphi_2$ so that we can make the
identification of $\varphi_2$ with $\varphi_2+\Delta$.  Moreover, this
shift symmetry has to hold at any order in the string loop expansion
and with any number of external states (``punctures'').  String loop
calculations are done with an Euclidean worldsheet, so the path
integral integrand is $\exp(-\pi^
{-1}\int{\cal L})$.  The field
$\varphi_2$ enters the Lagrangian $\cal L$ as ${\cal
L}_2=-\ft12\bar\partial\varphi_2\partial\varphi_2-hT_2$ (plus the
${\widetilde T}_2$, $W$ and ${\widetilde W}$ current terms, with which we
are not concerned at this point), where
$$
{\cal T}_2=-\ft12(\partial\varphi_2)^
2-Q_2\partial^
2\varphi_2.\eqno(50)
$$
By integrating the derivatives in the background charge term over onto
the spin-2 gauge field $h$, and noting that in light-cone gauge the
Ricci scalar $R$ is given by $R=2\partial^
2h$, we see that the
background charge term in (50) may be reinterpreted [6] as a dilaton
coupling ${\cal L}_{\rm dil}=R\Phi$ with $\Phi=\ft12Q_2\varphi_2$.
Thus, we have to ensure that the shift term
$\exp(-(2\pi)^
{-1}Q_2\Delta\int R)$ in the path integral is equal to
unity.  The quantity $(4\pi)^
{-1}\int R=\chi$ is the worldsheet Euler
number and so takes integer values.  (For unpunctured worldsheets,
$\chi$ takes even values, but punctures can make it odd.)   We must
also have $Q_2=-\im B_2$ imaginary (so we require $D\ge25$ here).
Thus, for compactification of the $\varphi_2$ direction, there will be
a symmetry under shifts of $\varphi_2$ by $\Delta=\pi/B_2$.

     The above discussion shows that we may compactify the $\varphi_2$
direction on a circle of circumference $\pi/B_2$ in the case
$D\ge25$.  Under these circumstances, the momentum $p_2$ can take the
values
$$
p_2=2nB_2, \qquad n\in\Z.\eqno(51)
$$
This implies that the $D$-dimensional mass formula following from (34)
upon compactification becomes
$$
{\cal M}^
2=-p_\mu
p^
\mu=\ft1{12}(26-D)+\ft1{12}(2n-1)^
2(D-\ft{49}2).\eqno(52)
$$
 From this it is clear that there are no massless states at this level
in the compactified theory:  the allowed values of $p_2$ cause ${\cal
M}^
2$ to ``jump over'' the value zero.  This accords with what we
found in our norm calculations in section 3, where there was no
longitudinal null state at level 1.  Since this is the level that,
when combined with level 1 in the right-moving sector, gives the
graviton in ordinary string theory, we come to the conclusion that
$W_3$ strings do not contain a massless graviton.  It seems likely
that the impossibility of massless states will continue at higher
levels as well.  This conclusion is in contrast to the general
arguments given in [7], which suggested that there should be
higher-spin massless states.  The arguments of [7] were based upon the
observation that the intercept for the $L_0$ mass-shell constraint is
$(-4)$, rather than the $(-1)$ of ordinary string theory, thus
suggesting a downward shift of the $W_3$ (mass)$^
2$ eigenvalues relative to
ordinary string theory.  However, this observation overlooks the fact
that $W_3$ strings require background charges, which more than overcome
the negative displacement of the intercept.  In fact, as we have seen,
the relative downward shift of $-3$ in the $L_0$ intercept is
counteracted by an upward shift of $\ft12Q_1^
2=3+\ft1{16}$ from the
fixed background charge in the $\varphi_1$ direction.

      For the level-1 states, the norm calculation in section 3 led to
the conclusion, supported by our above discussion for the compactified
theory at $D\ge25$, that it is misleading to view the
left-hand side of the mass-shell condition (34) as the ``$({\rm mass
})^
2$ operator.'' In fact, it would seem that a better indication of
mass is provided by the quantity $-{\bar p}^
a p_a$, which, for level-1
takes the value
$$
-{\bar p}^
a p_a=\ft18,\eqno(53)
$$
independent of $D$.  A similar calculation at level-1 in ordinary string
theory, with criticality achieved by taking $D$ spacetime coordinates with an
appropriate background charge, would yield the results $-\ph^
a\ph_a=\ft1{12}(26-D)$, and $-{\bar p}^
a p_a=0$.  Again,
it would seem that $-{\bar p}^ a p_a$ provides the better measure of mass:
In this case one finds by calculating the norm of the level-1 states that for
any $D$ there is always a null state, corresponding to a gauge degree of
freedom, and so it seems appropriate to view the level-1 states of ordinary
critical string theory as comprising a massless vector for all $D$.

     If one chooses a value for $D$ that is $\le24$, then the
compactification procedure that we have just discussed cannot be
implemented, since the background charge $Q_2$ is then real and so the
integrand of the functional integral will not exhibit any periodicity
under shifts in $\varphi_2$.  Thus, under these circumstances one cannot
use compactification to circumvent the difficulties of defining mass in
the $(D+1)$-dimensional spacetime with background charge. However, having
seen in the case of the spin-1 states that the most appropriate mass
operator appears to be $-{\bar p}^
a p_a$ rather than $-\ph^
a\ph_a$, one
might argue that the former should also be regarded as defining the mass
for states at all levels.  In particular, according to this viewpoint,
the mass for the ``tachyon'' state (15) (with $p_1=\im Q_1$) would be
given not by (31), which suggests that the ``tachyon'' would be massless
in $D=2$, but rather by $-{\bar p}^
a p_a$, which would give $({\rm
mass})^
2=-\ft{15}8$ for all $D$.  Since for the tachyon one does not have
gauge invariance as a guide, nor for $D\le24$ can one compactify to a
Minkowski spacetime with unbroken Lorentz symmetry, there is an inherent
ambiguity in the definition of the tachyon's mass.  It should be noted
that our preference for $-{\bar p}^
a p_a$ differs from the customary
choice in some of the recent literature on two-dimensional gravity,
where $-\ph^
a\ph_a$ is taken to be the $({\rm mass})^
2$ operator for the
tachyon, giving the appearance of a ``massless tachyon'' in $d=2$
spacetime.

\bigskip
\noindent{\bf 5. Discussion}

     In this paper, we have examined the spectrum of physical states for
critical $W_3$ string theories, and shown how the intrinsically
``non-stringy'' coordinate $\varphi_1$ appearing in the matter
realisations (1$a,b$) of the $W_3$ algebra is effectively ``frozen'' at
the quantum level by the $W$ constraint,  so that the momentum in the
$\varphi_1$ direction takes the fixed imaginary value $\im
Q_1$.  Moreover, physical states cannot contain any creation operators in
the $\varphi_1$ direction.  The remaining ($L_n$) constraints on physical
states give rise to a spectrum that is rather similar to that for a
critical string propagating in $d\ne26$ dimensions with a background
charge.  The main difference is that there appear to be no massless
states in the $W_3$ string spectrum.

     In deriving our results for the freezing of the $\varphi_1$
coordinate, we made the assumption that $\varphi_1$ is a spacelike
dimension.  One might wonder whether one could instead choose to take
$\varphi_1$ to be timelike, expecting that it will be frozen by the $W$
constraint in any case.  A consequence of this choice is that the
momentum shift for $p_1$, which follows from the analogue of the
hermiticity requirement (8),  becomes real rather than imaginary.
Because of this, the polynomial equations for $p_1$ that follow from
imposing the $W$ constraint on physical states admit more solutions.
For example, at level 0, the $W$ constraint now implies that
$p_1$ may be any root of the cubic equation $(p_1+Q_1)(p_1+\ft67 Q_1)
(p_1+ \ft87 Q_1)=0$ [1,8].  One also finds in that case that the discussion of
section 2 on the absence of $\alpha^
1_{-n}$ creation operators in physical
states is modified.  Since for timelike $\varphi_1$ the $p^1$ momentum
shift is real, the characteristic equations for the $W_0$ constraint
now do allow compatible solutions.  For example, at level 1 there are
now states compatible with the $W_0$ constraint that involve the
$\alpha^
1_{-1}$ oscillator.  However, when one calculates the norms of
these states, one finds that some of them are negative.  Thus even
though it is ``frozen'' by the $W$ constraint, a timelike $\varphi_1$
would give rise to a non-unitary theory.
\bigskip\bigskip
     \centerline{\bf ACKNOWLEDGMENTS}
\bigskip

     We are grateful to E. Bergshoeff for discussions. C.N.P. and K.S.S.
wish to thank the International Center for Theoretical Physics for
hospitality.
\bigskip\bigskip

\singlespace
\centerline{\bf REFERENCES}
\frenchspacing
\bigskip

\item{[1]}C.N. Pope, L.J. Romans and K.S. Stelle, ``On $W_3$  strings,''
preprint CERN-TH.6171/91, {\sl Phys.\  Lett.}\ {\bf B} (in press).

\item{[2]}C.N. Pope, L.J. Romans and K.S. Stelle, ``Anomaly-free $W_3$ gravity
and critical $W_3$ strings,'' preprint CERN-TH.6171/91, {\sl Phys.\  Lett.}\
{\bf B} (in press).

\item{[3]}J. Thierry-Mieg, {\sl Phys.\ Lett.}\  {\bf 197B} (1987) 368.

\item{[4]}V.A.\ Fateev and A.\ Zamolodchikov, {\sl Nucl.\  Phys.}\  {\bf
B280} (1987) 644;\nl
V.A.\ Fateev and S.\ Lukyanov,  {\sl Int.\ J.\ Mod.\  Phys.}\ {\bf
A3} (1988) 507.

\item{[5]}L.J.  Romans, {\sl Nucl.\  Phys.}\ {\bf B352} (1991) 829.

\item{[6]}I. Antoniadis, C. Bachas, J. Ellis and D.V. Nanopoulos, {\sl Nucl.\
Phys.}\ {\bf B328} (1989) 117.

\item{[7]}A. Bilal and J.L. Gervais, {\sl Nucl.\ Phys.}\ {\bf B326} (1989) 222.

\item{[8]}S.R.\ Das, A.\ Dhar and S.K.\ Rama, ``Physical properties of $W$
gravities and $W$ strings,'' preprint, TIFR/TH/91-11;\nl ``Physical states
and scaling properties of $W$ gravities and $W$ strings,''\nl
preprint, TIFR/TH/91-20.

\bye